\documentclass[aps,prb,showpacs,twocolumn,amsmath,amssymb,superscriptaddress,floatfix]{revtex4}

\usepackage{epsfig}
\usepackage{epstopdf}
\usepackage{graphicx}
\usepackage{grffile}
\usepackage{bm}
\usepackage{sidecap}
\usepackage{mathptmx}
\usepackage{txfonts}
\usepackage[colorlinks,citecolor=blue,linkcolor=Magenta]{hyperref}
\usepackage[usenames,dvipsnames,svgnames]{xcolor}
\usepackage{array}
\usepackage{float}

\usepackage{lipsum}

\newcommand{\tobedeleted}[1]{\textcolor{green}{#1}}
\renewcommand{\tobedeleted}[1]{\relax}

\begin{document}
\renewcommand{\thefigure}{\arabic{figure}}
\def\be{\begin{equation}}
\def\ee{\end{equation}}
\def\ber{\begin{eqnarray}}
\def\eer{\end{eqnarray}}

\def\kv{{\bf k}}
\def\qv{{\bf q}}
\def\pv{{\bf p}}
\def\sigmav{{\bf \sigma}}
\def\tauv{{\bf \tau}}
\newcommand{\h}[1]{{\hat {#1}}}
\newcommand{\hdg}[1]{{\hat {#1}^\dagger}}
\newcommand{\bra}[1]{\left\langle{#1}\right|}
\newcommand{\ket}[1]{\left|{#1}\right\rangle}

\title{ Effect of Rashba splitting on RKKY interaction in topological insulator thin films }
\date{\today}

\author {Mahroo Shiranzaei}
\affiliation{School of Physics, Damghan University, P.O. Box 36716-41167, Damghan, Iran}
\author{Fariborz Parhizgar}
\affiliation{School of Physics, Institute for Research in
Fundamental Sciences (IPM), Tehran 19395-5531, Iran}
\author{Hosein Cheraghchi}\email{cheraghchi@du.ac.ir}
\affiliation{School of Physics, Damghan University, P.O. Box 36716-41167, Damghan, Iran}

\begin{abstract}
In this work we have investigated the effect of Rashba splitting on the RKKY interaction in TI thin film both at finite and zero chemical potential. We find that the spin susceptibility of Rashba materials including TI thin film is strongly dependent on the direction of distance vector. Moreover, we find another term in the off-diagonal terms of the spin-susceptibility tensor which in contrast to the well-known DM-like term is symmetric. We show how one can tune the RKKY interaction by using electric field applied perpendicularly to the surface plane and by small chemical doping giving rise to enhance the RKKY term, drastically. We have presented our results for two different situations, namely inter-surface pairing of magnetic impurities as well as intra-surface one. The behavior of these two situations is completely different which we describe it by mapping the density of states of each surface on the band dispersion.
\end{abstract}
\maketitle

\section{Introduction}
Among different types of magnetic interaction detected in materials, Ruderman-Kittle-Kasuya-Yosida (RKKY)\cite{Ruderman, Kasuya, Yosida} mechanism, an indirect exchange interaction between two magnetic adatoms via host itinerant electrons, is one of the main reasons of coupling between magnetic impurities. This interaction is proportional to the spin susceptibility of the host material and so gives the spin information of the system.\cite{parhizgar-mos2, parhizgar-spg} Depending on the spin structure of the host material, different types of couplings can occur between magnetic adatoms via the RKKY interaction. While in spin-degenerate systems, such as graphene,\cite{sherafati, annica, parhizgar-bg, szalowski} two localized magnetic impurities couple to each other in the form of isotropic collinear Heisenberg-like term, the anisotropic collinear Ising-like term with different coefficients in different spin-directions can be appeared in spin-polarized systems.\cite{parhizgar-spg, valizadeh} Moreover in materials with Rashba spin-orbit coupling \cite{imamura, TI-RKKY, pesin} as well as materials with spin-valley coupling, \cite{parhizgar-mos2, zareSi, jelena-CNT} it has been shown that twisting RKKY interaction is possible by the anti-symmetric non-collinear Dzyaloshinskii-Moria-like term.\cite{dm1,dm2} In general, the RKKY is a long-ranged interaction, (it decays with $R^{-D}$, $D$ the dimension of the system) which oscillates with respect to the distance between impurities and electron's Fermi wave-vector.
 Fascinating feature of this well-known mechanism is measurable in experimental observations by angle-resolved photoemission spectroscopy (ARPES) and scanning tunneling microscopy (STM) in the study of magneto-transport and single-atomic magnetometry.\cite{prl91116601, nature497, nature187, zareSi}

Moreover, the RKKY interaction can be in charge of diverse magnetic phases and ordering in metals and semiconductors\cite{sherafati, szalowski1, power} such as ferromagnetic and anti-ferromagnetic \cite{Hwang, Priour, Matsukura, Ko-2011, Ohno-1998} as well as spin glass\cite{prb5115250, prb36492} and spiral phases.\cite{zareSi, paaske}
Recently, quantum anomalous Hall effect (QAHE) have been predicted theoretically\cite{QAHE-theor} and experimentally realized\cite{QAHE-exp1, QAHE-exp2, QAHE-exp3} in magnetically doped three-dimensional ($3$D) topological insulators. Since such experiments need the ferromagnetic coupling of magnetic adatoms, it brings intensive attentions to the mechanism of the coupling among magnetic impurities in this class of materials. Although the RKKY interaction (and more precisely its zero chemical potential version known as Van-Vleck mechanism) is thought to be the main mechanism of this coupling,\cite{Zhang-PRL} such theory is still under debate.\cite{peixoto}

$3$D topological insulators (TI), systems with gapped bulk states and gapless surface states protected by time reversal symmetry (TRS), are a novel kind of materials that have been subject of several researches during past few years.\cite{Hasan-rev, Qi-rev, Moore} An important branch of these topological insulators is Bismuth-based structures, for instance Bi$_2$Se$_3$ and Bi$_2$Te$_3$, which made of Van-der-Waals interacting layers known as Quintuple Layers (QL).\cite{nature438}
For thicknesses above $6$QLs, the Bismuth-based materials become topological insulator with gapless surface states\cite{crossover} which have isotropic Dirac-type band dispersion presenting with an effective chiral Hamiltonian arising from pure Rashba-type spin-orbit coupling. Combination of the pure Rashba Hamiltonian with being in the category of Dirac materials\cite{Wehling-rev} makes TIs a promising candidate for spintronic and electronic applications.\cite{pesin-nature} Since the bulk band gap of these $3$D systems are not enough large, in practice, the bulk states usually play a severe role in experiments and so it is more favorable to use thin version of these structures in order to reduce the effect of their bulk states.
 It has been experimentally shown that for $5$QLs thickness and less, the states of different surfaces of TI thin film would be hybridized. Although these ultra thin films are not $3$D topological insulators with gapless states, they can share other interesting features such as another topological phase transition from quantum spin Hall insulator to a normal insulator,\cite{Linder, Liu-2010, Shan-NJP, Zhang-PRL} time reversal topological superconductivity\cite{fariborz-TSC} and band tunability by applying perpendicular electric\cite{fariborz-annica, fariborz-TSC, Zhang-PRL} or in-plane magnetic field.\cite{fariborz-opt}
Furthermore, magnetic topological insulators and their thin version\cite{Sessi-2014, Wray-2010} become of much importance since the ordered magnetic impurities on the surface of TI can create a magnetic field and open a gap in the band dispersion which has been observed experimentally.\cite{Wray-2010} Such intrinsic  ferromagnetism can result in QAHE when the Fermi energy lies within the gap of the system.
The RKKY interaction in the Rashba materials such as TIs have been explored extensively.\cite{imamura, prl102, TI-RKKY, pesin, Efimkin, Loss-SC} Existence of the strong Rashba spin-orbit coupling in these materials makes the RKKY interaction to have a rich physics that includes a DM-like term \cite{TI-RKKY} and can result in different magnetic phases such as ferromagnetic, paramagnetic and spin-glass.\cite{pesin} In addition, such interaction on the surface of TI has been investigated when a superconductor presents in the proximity of TI.\cite{Loss-SC} Since the magnetic impurities ordered perpendicular to the surface of TI can produce a gap on the surface state, the RKKY interaction together with such gap has been investigated self-consistently.\cite{Efimkin} While all these theoretical investigations have been done for a thick $3D$ TI, the experimental realization of QAHE in TI thin films makes it essential to investigate the RKKY interaction in thin version of TIs where two surfaces be hybridized to each other.\cite{Zhang-PRL}

In this work, we have investigated the spin susceptibility of TI thin film and so the RKKY interaction both at zero and finite doping. In contrast to the most of previous works on TIs, we found strong spatial anisotropy of the RKKY interaction with respect to the direction of the connecting line between impurities when one or both impurities have an in-plane spin-component projected on the surface of TI.\cite{Efimkin} We tried to explore the effect of parameters such as chemical potential, tunnelling strength between surfaces and applied biased electric field on the RKKY interaction. The last one has the benefit that one can tune the RKKY interaction and as a result the magnetic properties, by using an electric field. We describe our findings by means of contribution of the top and bottom surface states in the band dispersion.
The organization of the paper is as follow: In section \ref{theory} we introduce the theory of the work starting with the model Hamiltonian. In this part, we present contribution of the top and bottom surfaces in the band dispersion separately which is so important to describe our results. Next, we report our method for calculating the RKKY interaction by using the real space Green's function. To guarantee fluency, we have presented some details of calculations and also analytic results for the real space Green's functions in the appendix \ref{app}.
Section \ref{result} presents our results where we discuss the RKKY interaction between impurities on the same and different surfaces. We have summarized and concluded our results in section \ref{conclusion}.

\section{Theory}\label{theory}

\subsection{Model Hamiltonian}\label{Hamil}

The surface states of the TI thin film around the $\Gamma$ point can be described by the two-dimensional effective Hamiltonian  \cite{crossover, hamiltonian, effective}
\begin{equation}\label{eq:1}
   H_0(k)=-D\,k^2\sigma_0\otimes\tau_0+[\hbar v_{F} (\sigma\times \textbf{k})\cdot \hat z + V\sigma_0]\otimes\tau_z+\Delta\sigma_0\otimes\tau_x,
\end{equation}
where ${\bf \sigma}$, ${\bf \tau}$ are Pauli matrices in spin and surface space respectively, $\textbf{k}=(k_x,k_y)$ represents the wave-vector of surface state's electrons and $v_F$ is their Fermi velocity. The term with coefficient $D$ refers to the particle-hole asymmetry in the system and $V$ shows the potential difference between surfaces which can be caused by the effect of substrate or an external electric field applied perpendicularly to the surfaces. The last term in the above equation shows the tunneling between different surfaces and in general it is of the form $ \Delta-\Delta_1 k^2 $ where the $\Delta_1$ term can result in a topological phase transition in the system with potentials lower than a critical value $V<\hbar v_F \sqrt{\Delta/\Delta_1}$ for special thin films in which $\Delta \; \Delta_1>0$.\cite{effective, prl115} Restricting ourselves to the terms upto the first order in $k$, the energy dispersion would be obtained as
\begin{equation}\label{eq:2}
  E(k)=\pm\sqrt{(\hbar \, v_{F} \,k +\mp V)^2+\Delta^2},
\end{equation}
where the sign $\pm$ before the root square is related to the conduction (C) and valance (V) bands and the sign $\mp $ before parameter $V$ refers to the different branches $(1,2)$ in each of (C,V) bands that has been separated as a result of the applied potential $V$ known as Rashba splitting in the band dispersion.

\begin{figure}[!h]
  \includegraphics[scale=0.62]{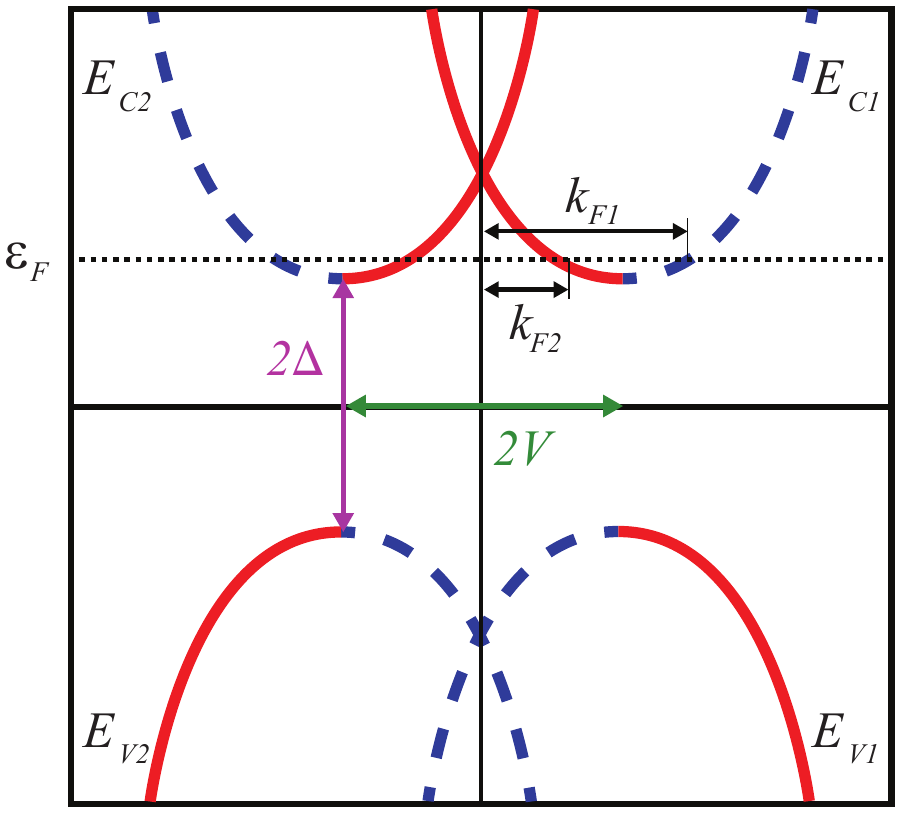}\\
  \caption{\small{(Color online) Schematic figure shows the dispersion of TI thin film and illustrates two Fermi wave-vectors, $k_{F1,2}$, and the Fermi energy $\varepsilon_F$ by the dotted line. }}\label{fig.schematic}
\end{figure}

A schematic figure of these band dispersions has been depicted in Fig.\ref{fig.schematic}. In this figure the horizontal dotted line shows the chemical potential, which together with the applied potential $V$ are tunable parameters of the system. As a result of Rashba splitting, two different Fermi wave-vectors $k_{F1,2}=(\sqrt{\varepsilon_F^2-\Delta^2}\pm V)/\hbar v_F$ appear in the system

 Also in this figure, the red solid lines (the blue dashed lines) show the criteria that the band dispersion comes mostly from the top (bottom) surface.\cite{effective} This can be better understood by looking at the Green's function of the system where the local density of states (DOS) of the top surface can be studied separately from the bottom surface and its poles represent the band dispersion. By using $G_0(k,\varepsilon)=(\varepsilon-H_0(k))^{-1}$, we have

\begin{equation}\label{eq:3}
G_0(k,\varepsilon)=\begin{bmatrix}
  g_{t\uparrow t\uparrow} & g_{t\uparrow t\downarrow} & g_{t\uparrow b\uparrow} & g_{t\uparrow b\downarrow} \\
  g_{t\uparrow t\downarrow}^* & g_{t\uparrow t\uparrow} & g_{t\uparrow b\downarrow}^* & g_{t\uparrow b\uparrow} \\
 g _{t\uparrow b\uparrow} & g_{t\uparrow b\downarrow} & g_{b\uparrow b\uparrow} & g_{b\uparrow b\downarrow} \\
  g_{t\uparrow b\downarrow}^* & g_{t\uparrow b\uparrow} & g_{b\uparrow b\downarrow}^* & g_{b\uparrow b\uparrow}
\end{bmatrix},
\end{equation}
where, $ t (b) $ and $ \uparrow (\downarrow) $ refer to the top (bottom) surface and spin up (down) respectively. In addition, similarities between the components have been considered in this matrix. Focusing on just the diagonal elements of the Green's function which are required for calculation of the DOS, we would have

\begin{align}\label{eq:4}
g_{t\uparrow t\uparrow}(k, \varepsilon)=\frac{A^+}{(\varepsilon-E_{V1})}+\frac{A^-}{(\varepsilon-E_{C1})}+\frac{B^-}{(\varepsilon-E_{V2})}+\frac{B^+}{(\varepsilon-E_{C2})}\nonumber\\
g_{b\uparrow b\uparrow}(k, \varepsilon)=\frac{A^-}{(\varepsilon-E_{V1})}+\frac{A^+}{(\varepsilon-E_{C1})}+\frac{B^+}{(\varepsilon-E_{V2})}+\frac{B^-}{(\varepsilon-E_{C2})}\nonumber,\\
\end{align}
where coefficients $ A^{\pm} $ and $ B^{\pm} $ are functions of $ k $, $ \Delta $ and $ V $ as the below

\begin{equation}\label{eq:5}
\begin{split}
  A^{\pm}=\frac{\sqrt{\Delta ^2+(k-V)^2} \pm (k-V)}{4 \sqrt{\Delta ^2+(k-V)^2}}, \;
  B^{\pm}=\frac{\sqrt{\Delta ^2+(k+V)^2} \pm (k+V)}{4 \sqrt{\Delta ^2+(k+V)^2}}
\end{split}.
\end{equation}

Using Dos $\propto \frac{-1}{\pi}\sum_k Im[G(k,\varepsilon)]$ and the fact that the imaginary part of the Green's function is peaked on the poles of Eq.\eqref{eq:4} as $\delta(\varepsilon-E(k))$, one can interpret the coefficients $A^{\pm}, B^{\pm}$ as the weight coefficients of the DOS on different band dispersions.

\begin{figure}[!h]
  \includegraphics[scale=0.32]{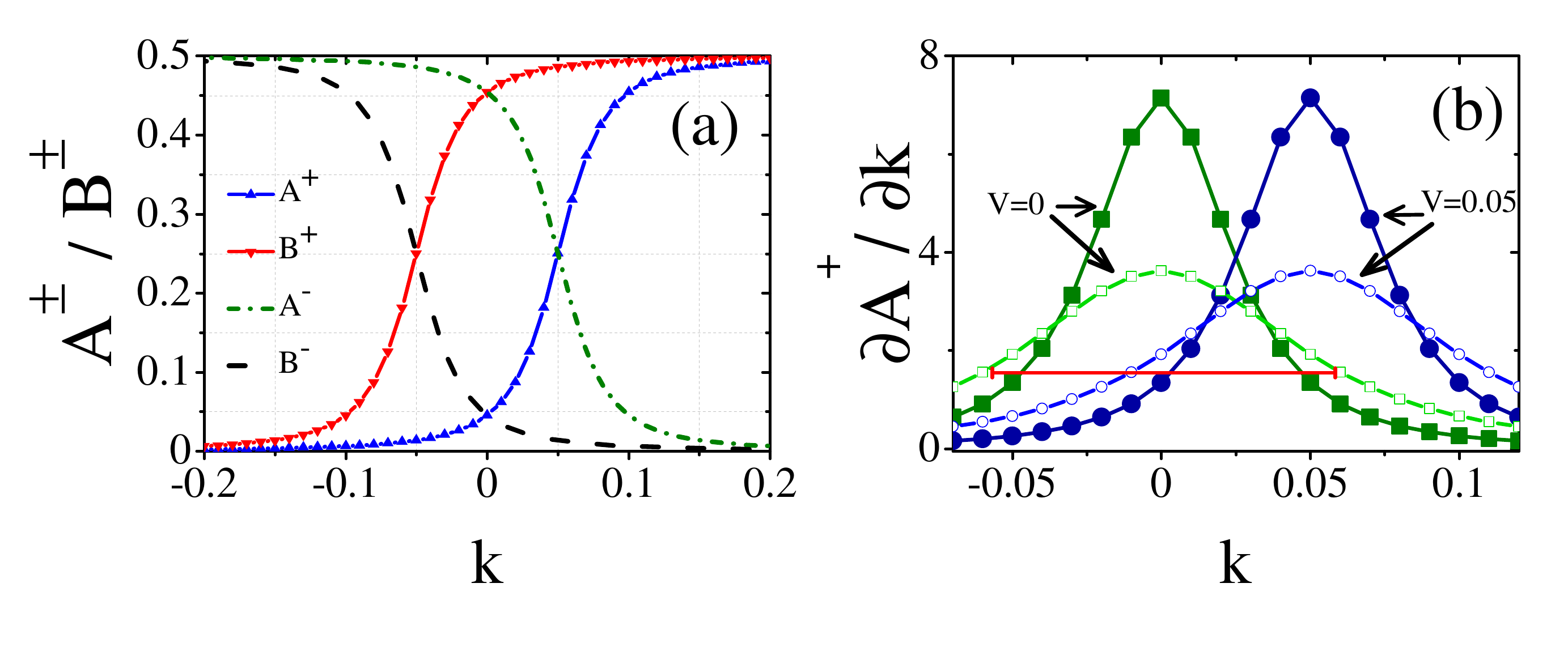}\\
  \caption{\small{(Color online) (a) The weight coefficients $A^{\pm}$ and $B^{\pm}$, as a function of $k$ for $V=0.05$ eV and $\Delta=0.035 $ eV. (b) partial derivative of coefficient $A^+$ with respect to $k$ for two different potentials $V=0,0.05$eV and two different gap sizes $\Delta=0.035$eV (solid symbols) and $\Delta=0.069$eV (hollow symbols).}}\label{fig.coefficient}
\end{figure}

Fig.\ref{fig.coefficient} (a) shows the behaviour of the weight coefficients $A^{\pm},B^{\pm}$ as a function of $k$. As shown in this figure, the weight of the conduction band $E_{C1}$ ($E_{C2}$) at large positive (negative) $k$ is dominated by the bottom (top) surface. At $k=0$ the dominant contribution of both conduction (valance) bands are originated from the top (bottom) surface state. 
It should be mentioned that surface states hybridization emerges around the band edges where group velocity of carriers is zero. The region where surface states are hybrid with each other depends only on the tunnelling between two surfaces $\Delta$ and not on the applied bias $V$. Figure \ref{fig.coefficient} (b) shows the k-derivative of the coefficient $\frac{\partial A^{+}}{\partial k}$  for different biased potentials $V=0,0.05$eV and two different gap sizes $\Delta=0.035$eV (solid symbols) and $0.069$eV (hollow symbols). These diagrams are peaked functions of $k$ with the widths proportional to $\delta k \propto \Delta$. The bias voltage $V$ can just change the position of these peaks with no effect on their widths.

\begin{figure}[!h]
  \includegraphics[width=1\linewidth]{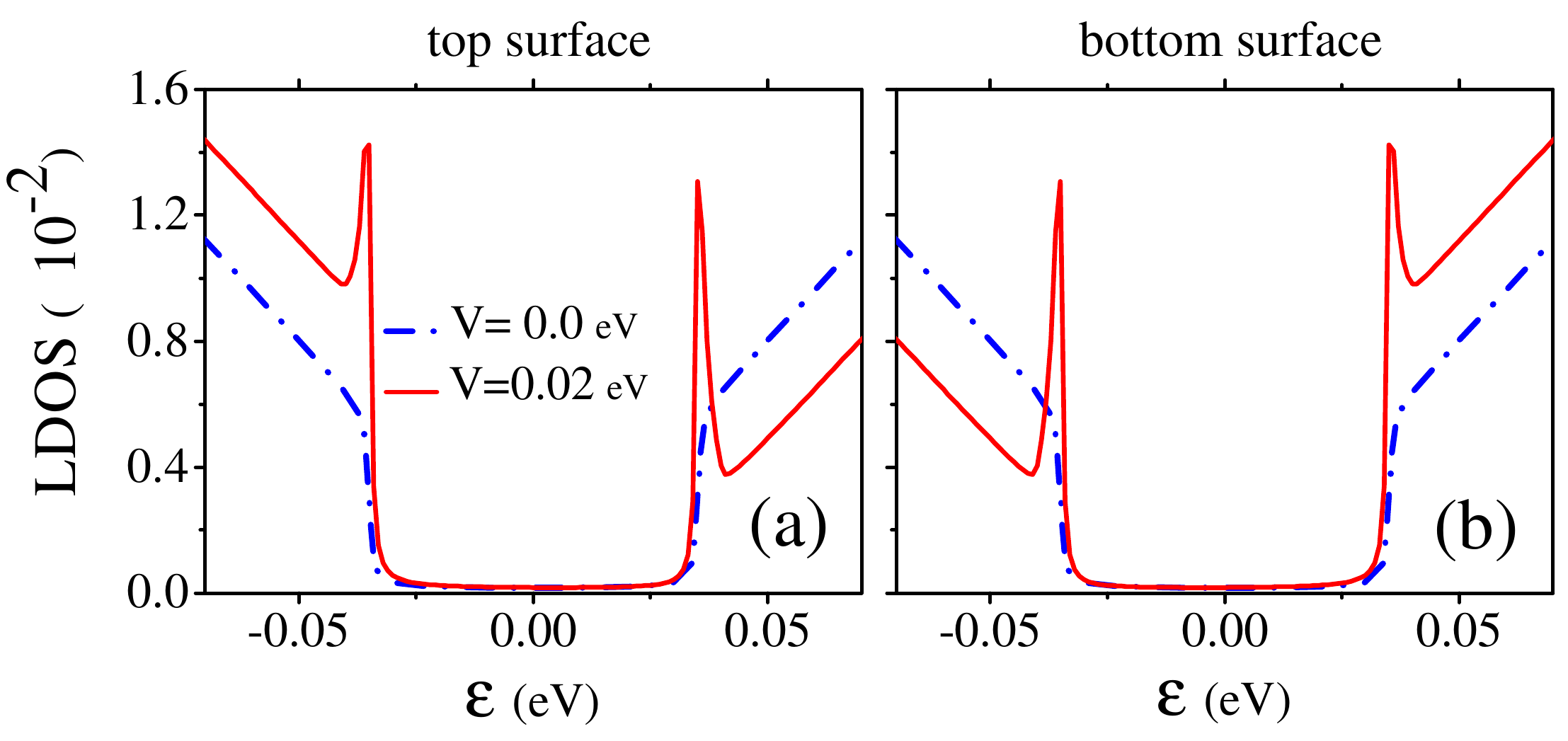}
  \caption{\small{(Color online) Illustration of the density of states for unperturbed system for two different values of voltage  $V=0$ eV and $V=0.02$ eV for (a) the top and (b) the bottom surfaces.}}\label{fig.LDOS}
\end{figure}

In addition to the weight coefficients for the band dispersion which we will use them in the result section, one can calculate the density of states (DOS) for the top and bottom surface separately. Fig.\ref{fig.LDOS} shows the DOS of different surfaces for fixed tunnelling parameter $\Delta=0.035$ eV and two different values of voltage $V=0, \; 0.02$ eV. As one can see, Van-Hove singularities appear in the DOS due to Rashba splitting ($V\neq 0$) when the energy touches the boundaries of the gap $\varepsilon=\pm\Delta$.

\subsection{RKKY interaction}
By placing two magnetic impurities on the surfaces of TI thin film, the Hamiltonian would be modified to
\begin{align}\label{eq:6}
H=H_0(k)+J_c\sum_{i=1,2}S_i \cdot \hat{s({\bf r}_i)},
\end{align}
where $S_i$ shows the spin moment of the localized magnetic impurity, $\hat{s}({\bf r}_i)=\hbar/2 \sum_j \sigma_j \delta({\bf r}-{\bf r}_j)$ denotes the spin of itinerant electrons and $J_c$ displays the coupling between them.
By applying the second order perturbation theory, one can transform the interaction between magnetic impurities and itinerant electrons to an indirect exchange interaction between two magnetic impurities. Thus, the RKKY interaction would read as \cite{grosso, ruderman, coleman}
\begin{equation}\label{eq:7}
  H_{RKKY}^{\alpha \beta}=J_{c}^2\sum_{i,j}S_{i}^{\alpha}\chi_{ij}^{\alpha\beta}(r,r^\prime)S_{j}^{\beta},
\end{equation}
where $\chi_{ij}^{\alpha\beta}(r,r^\prime)$ is the spin susceptibility of the system and can be evaluated as
\begin{equation}\label{eq:8}
\begin{aligned}
  &\chi_{ij}^{\alpha\beta}(r,r^{\prime})= \\
  &\frac{-1}{2\pi}Im\int_{-\infty}^{\varepsilon_F} d\varepsilon \ Tr[\sigma_{i}\ G^{\alpha\beta}(r,r^{\prime},\varepsilon)\ \sigma_{j}\ G^{\beta\alpha}(r^{\prime},r,\varepsilon)].
  \end{aligned}
\end{equation}
Here, $\alpha$ and $\beta$ denote t/b surface, $(i,j)=(x, y, z)$ show different directions of magnetic moment's component, $\varepsilon_F$ refers to the Fermi energy and trace is taken over the spin degree of freedom.

In order to calculate the spin susceptibility Eq.\eqref{eq:8}, it is needed to calculate the unperturbed retarded Green's function in real space, $  G_0^{ret}(\varepsilon,R) $ which reads from the Green's function in k-space Eq.\eqref{eq:3} by taking Fourier transformation

\begin{equation}\label{eq:9}
\begin{split}
 G_0^{ret}(\varepsilon,R=r_1-r_2)=\frac{1}{\Omega_{BZ}}\int \ d^2k \ e^{i \, \vec{\textbf{k}} \, \cdot \, \vec{\textbf{R}}} \, G_0(k).
      \end{split}
\end{equation}
Such Green's function has a general form of

\begin{align}\label{eq:10}
&G_0^{ret}(\varepsilon,\pm R)= \nonumber
 \\& \begin{bmatrix}
    G_{tt} & \mp e^{-i \, \varphi_R} \, G'_{tt}  & \vdots & G_{tb} & \mp e^{-i \, \varphi_R} \, G'_{tb} \\
   \pm e^{i \, \varphi_R} \, G'_{tt} & G_{tt} & \vdots & \pm e^{i \, \varphi_R} \, G'_{tb} & G_{tb} \\
    \dots & \dots & \dots & \dots & \dots\\
    G_{tb} & \mp e^{-i \, \varphi_R} \, G'_{tb} & \vdots & G_{bb} & \mp e^{-i \, \varphi_R} \, G'_{bb} \\
    \pm e^{i \, \varphi_R} \, G'_{tb} & G_{tb} & \vdots & \pm e^{i \, \varphi_R} \, G'_{bb}& G_{bb} \\
  \end{bmatrix},
\end{align}
where the components of the Green's functions are given in Appendix \ref{app}.

The impurities can be both located on the same surface (intra-surface) as well as different surfaces(inter-surface). Although in the former case the position of impurities can be assumed to be both on the top or bottom surfaces but using the symmetry of layer inversion together with $V \rightarrow -V$, one can achieve the result of the bottom surface from the top one and so in the following, we discuss two configurations namely, impurities to be located on the top surface and impurities located on different surfaces. After some calculations, the RKKY Hamiltonian Eq.\eqref{eq:7} can be written as

\begin{align}\label{eq:11}
H_{RKKY}=J_H S_1\cdot S_2+J_I \tilde{S}_1\cdot \tilde{S}_2+{\bf J}_{DM}\cdot (\tilde{S}_1\times \tilde{S}_2)\nonumber\\+J_{xy}(\tilde{S}_{1x}\tilde{S}_{2y}+\tilde{S}_{1y}\tilde{S}_{2x}),
\end{align}
where the new spinors $\tilde{S}$ is defined as $\tilde{S}=(S_x\cos(\varphi),S_y\sin(\varphi),S_z)$, with $\varphi=\tan^{-1}(R_y/R_x)$ and also
the vector ${\bf J}_{DM}=J_{DM}(1,1,0)$ and $J_{xy}=J_I$. The details of the above terms including some analytic results can be found in appendix \ref{app}.

In conventional two-dimension materials with isotropic band dispersion, the RKKY interaction does not depend on the direction of ${\bf R}$, the distance vector between impurities. However, in systems with Rashba spin-orbit coupling, the spin of itinerant electrons is coupled to the wave-vector ${\bf k}$ and  an impurity with an in-plane magnetic moment would break the isotropy of the system, so the spin-response $\chi_{ij}({\bf R})$ depends on both magnitude and direction of the vector ${\bf R}$.\cite{Efimkin}

\begin{figure}[H]
  \includegraphics[width=1\linewidth]{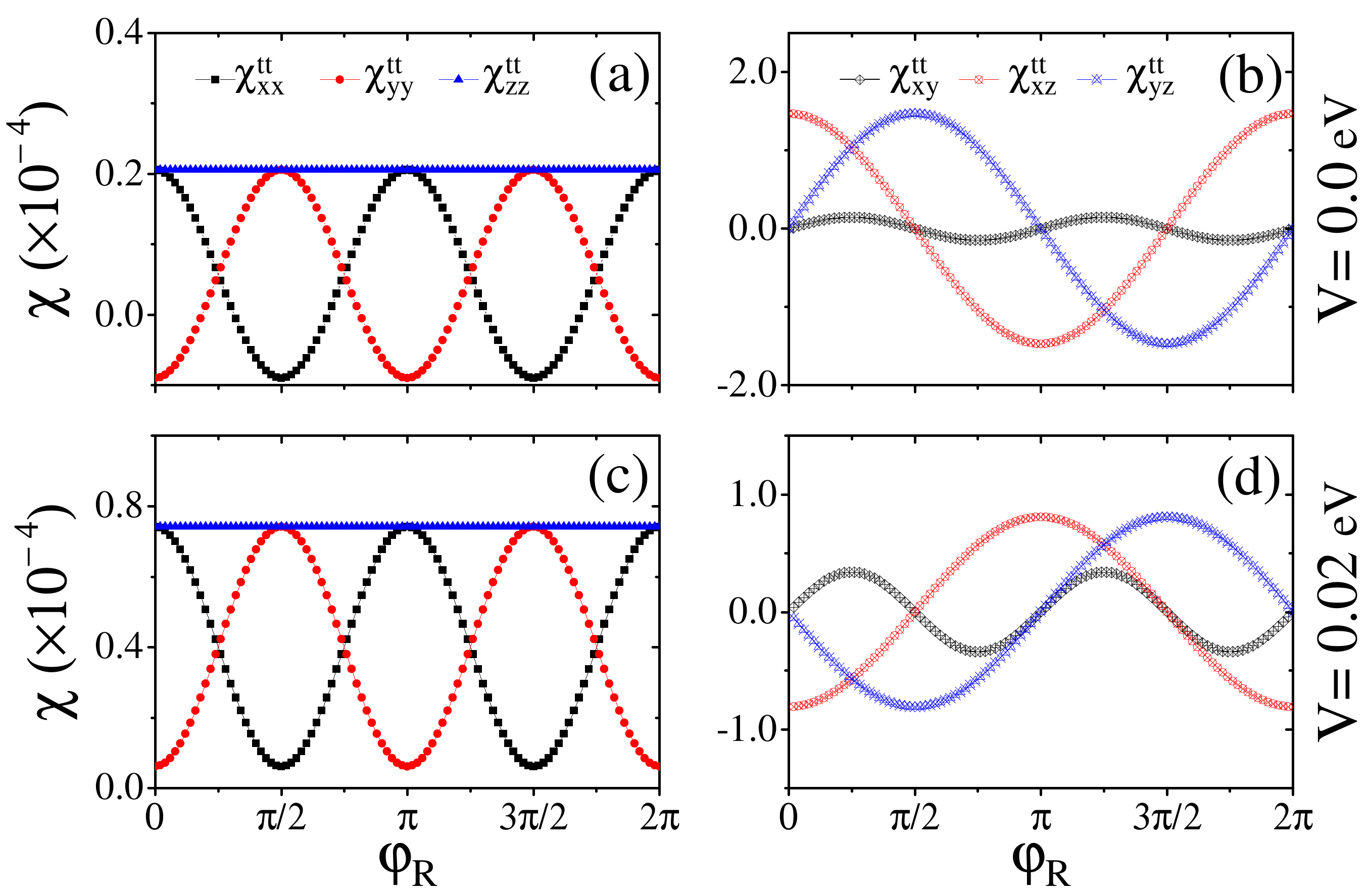} 
  \caption{\small{(Color online) The (a,c) diagonal, (b,d) off-diagonal components of susceptibility tensor, $\chi_{ij}^{\alpha \beta}$ as a function of polar angle $ \varphi_R $ are showed for inter surface case. All of them are scaled by $ (\frac{1}{\hbar^2 v_F^2 \Omega_{BZ}})^2 $. Here we set to $\Delta=0.035 $ eV, $\varepsilon_F=0.135 $ eV, $R=30$ nm, $ v_F=4.48\times 10^5  \frac{m}{s}$ and (a,b) $V=0$ eV, (c,d) $V=0.02$ eV.}}\label{fig.chiphiR}
\end{figure}

The RKKY interaction coefficients $J$s, introduced in Eq.\eqref{eq:11} for two considered configurations of impurities, intra-surface case ($tt$) or inter-surface case ($tb$), are defined as follow
\begin{align}\label{eq:12}
&J_H^{tt/tb}=-\frac{1}{\pi}\int_{-\infty}^{\varepsilon_F} d\varepsilon \; (G_{tt/tb}^2(\varepsilon, R)+G_{tt/tb}^{'2}(\varepsilon, R))\nonumber\\
&J_I^{tt/tb}=\frac{2}{\pi}\int_{-\infty}^{\varepsilon_F} d\varepsilon \; G_{tt/tb}^{'2}(\varepsilon, R)\nonumber\\
&J_{DM}^{tt/tb}=-\frac{2}{\pi}\int_{-\infty}^{\varepsilon_F} d\varepsilon \; G_{tt/tb}(\varepsilon, R) \; G'_{tt/tb}(\varepsilon, R).
\end{align}

The first term in Eq.\eqref{eq:11} is similar to the Heisenberg spin interaction which makes no difference between different spin-directions coupling. However, the second term couples the new spinors $\tilde{S}$ instead of $S$ and since $\tilde{S}$ depends on the angle $\varphi$, the $J_I$ couples spinors of two impurities which have different amplitudes in different directions. This interaction is similar to the Ising interaction. Both of these terms will result in collinear alignment of spinors $S_1$ and $S_2$. Moreover, due to the existence of Rashba spin-orbit coupling in TI thin film, symmetry of spin space is broken and so it is expected that the RKKY interaction would also have terms related to the off-diagonal components of the spin-susceptibility tensor.\cite{TI-RKKY,imamura} The third and forth terms of the above Hamiltonian are of this kind and contrarily with the first two terms, they can cause non-collinear twisted alignment between spinors of impurities. While the third term is anti-symmetric with respect to the spinors and resembles the Dzyaloshinskii-Moriya (DM) interaction, the last term is symmetric.
\begin{figure}[H]
  \includegraphics[width=0.85\linewidth]{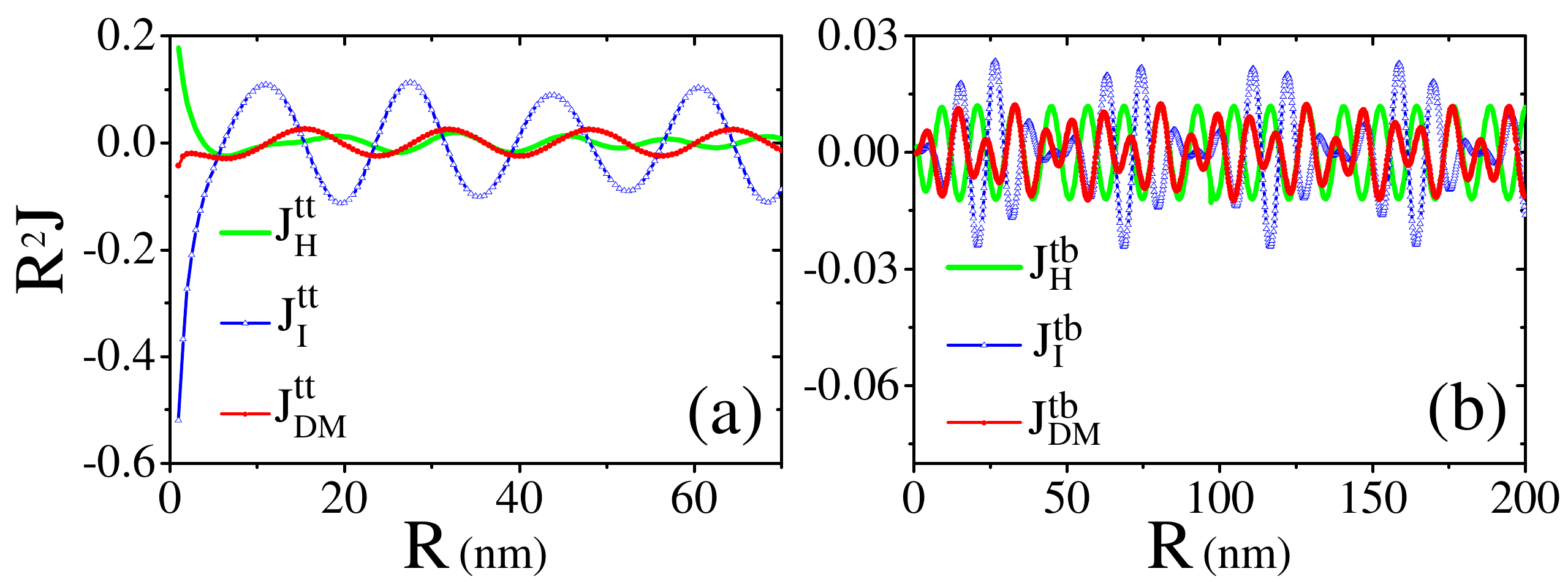}\\
  \caption{\small{(Color online) The RKKY interaction terms times $R^2$  ($R^2 J^{\alpha \beta}_i$ for $ i=H, I, DM $), as a function of the distance in unit of nm, scaled by $ (\frac{J_c}{\hbar^2 v_F^2 \Omega_{BZ}})^2$. Here we set  $\Delta=0.035$ eV, $V=0.02$ eV, $\varepsilon_F=0.085 $ eV and $v_{F}=4.48\times 10^5 \frac{m}{s}$. Panels (a), (b) refer to intra and inter-surface cases respectively.}}\label{fig.JR}
\end{figure}
\section{Results and discussions}\label{result}
In this section, we present our results for the RKKY interaction between two magnetic impurities located on the top surface (tt) or on two different surfaces (tb).
As we have shown in the previous section, the SOC in the topological insulator results in the angle-dependent of the RKKY interaction when magnetic moment of impurities have an in-plan component.
We start our result section by presenting this angle dependency of the spin susceptibility in Fig.\ref{fig.chiphiR}. In this figure, we have assumed both impurities to be located on the top surface and $\varepsilon_F=0.135$ eV, $\Delta=0.035$ eV and also we have considered two voltages, $V=0$ (panel (a,b)) and $V=0.02$ eV (panel (c,d)). We have plotted diagonal parts of the spin-susceptibility tensor in panels (a,c) where except $\chi_{zz}$, other terms oscillate with $2\varphi_R$. The off-diagonal parts depicted in panels (b,d) oscillate with $\varphi_R$ as expected from Eq.\eqref{eq:A3}.
Besides, by comparing the upper and lower panels it is specified how applying the voltage can drastically change sign and magnitude of the interaction terms.

The behaviour of the RKKY interaction terms are severely affected by distance between two magnetic impurities. In two-dimensional materials, they usually fall off with $R^{-2}$ and also oscillate as $\sim\sin(2k_FR)$, however for materials with several bands, a more complicated behavior is expected. In Fig.\ref{fig.JR}, we have plotted $ J_H, J_I, J_{DM} $, times $R^2$ and scaled by $ (\frac{J_c}{\hbar^2 v_F^2 \Omega_{BZ}})^2 $, in terms of distance $ R $ for  intra- (panel(a)) and inter- (panel(b)) surface case. As one can see in this figure, for the long range distances, all interaction terms decay as $R^{-2}$ as like as other two-dimensional structures.\cite{TI-RKKY, parhizgar-mos2} For the intra-surface pairing and in the short distance limit which plays a more prominent role at higher densities of impurities, the RKKY interaction has much higher values.

 In contrast to the intra-surface pairing between impurities, the RKKY interaction multiplied by $R^2$ behaves in a more strange way for the inter-surface pairing. First, it starts from nearly zero values at short distances and then it oscillates in a beating type pattern according to the existence of two different wave-vectors in the system.\cite{parhizgar-mos2, zareSi, valizadeh2} By looking at the weight coefficients in Eq.\eqref{eq:5}, one can see that for the top surface, $k_{F1}$ has a more prominent role rather than $k_{F2}$, and that's why it is seen a roughly monotonic oscillation for the RKKY interaction in the intra-surface case. However in the inter-surface case, both surfaces and as a result both $k_F$s become important and beating type occurs due to two different oscillations characterized by $k_{F1}$ and $k_{F2}$.

\begin{figure}[H]
  \includegraphics[width=1\linewidth]{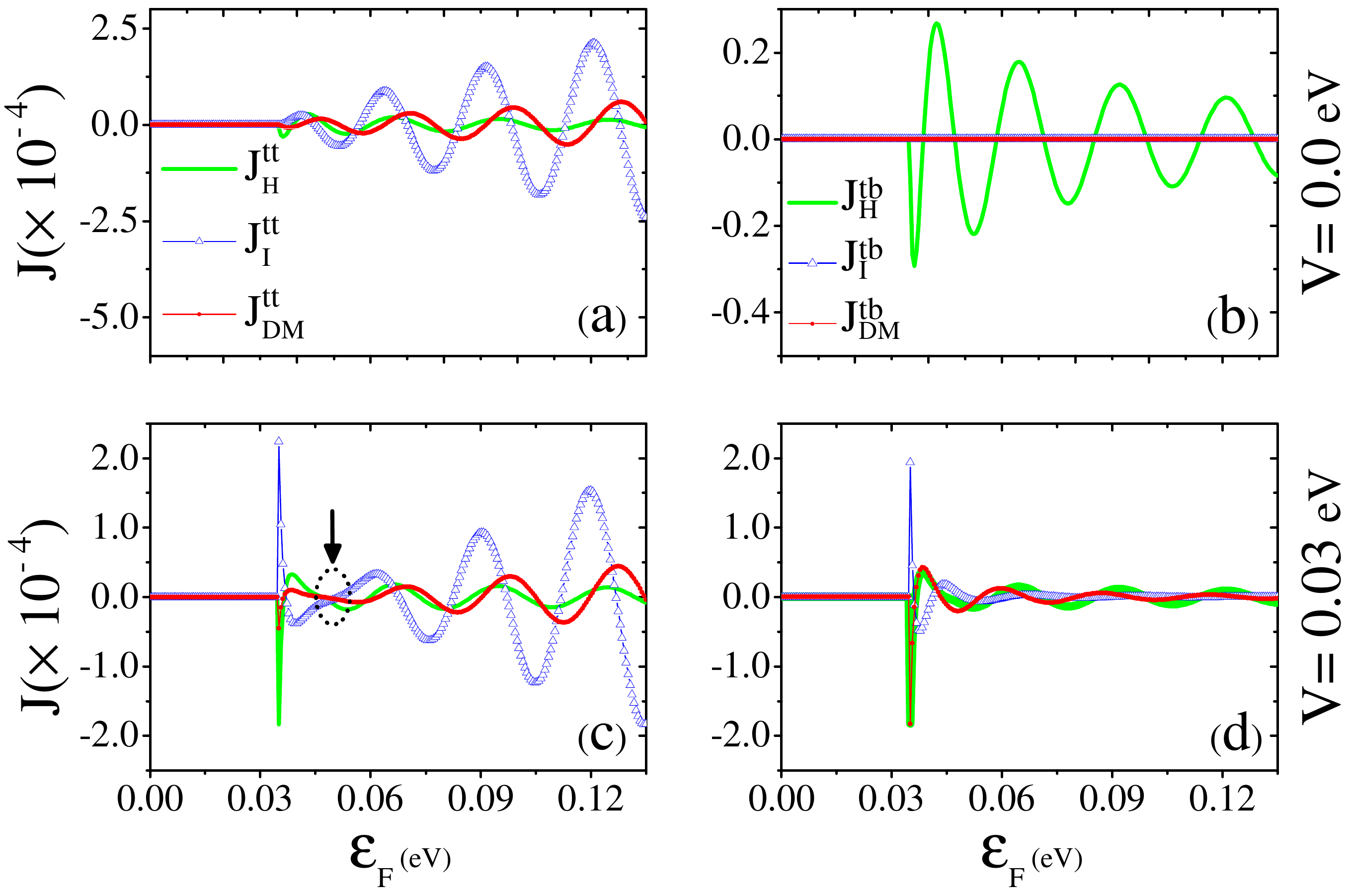}\\
  \caption{\small{(Color online) The intra- (a,c) and inter- (b,d) surface RKKY interaction couplings ($J_i^{\alpha \beta}$ for $ i=H, I, DM $), scaled by $ (\frac{J_c}{\hbar^2 v_F^2 \Omega_{BZ}})^2$ as a function of the Fermi energy in unit of eV. Here we set $\Delta=0.035$ eV, $R=30$nm, $v_{F}=4.48\times 10^5 \ \frac{m}{s}$, $V=0$ eV for panel (a,b) and $V=0.03$ eV for panel (c,d).}}\label{fig.JEf}
\end{figure}


Fig.\ref{fig.JEf} shows the effect of the Fermi energy on the RKKY interaction terms for intra-surface (a,c) and inter-surface (b,d) cases. Here, we chose $R=30$ nm, $\Delta=0.035$ eV and also $V=0$ eV in panels (a,b) and $V=0.05$ eV in panels (c,d).
As shown in these figures, for the Fermi energy inside the gap, $\varepsilon_F<\Delta$, all types of interactions are nearly zero according to insulating nature of the material, however they are not exactly zero and have small values which refer to the Van-Vleck mechanism.\cite{prl115, science329}
By comparing the insets of panel (a) and (c) in Fig.\ref{fig.JEf}, it is observed that all the interaction terms change with the potential $V$ and none of them can be neglected in favour of another.
At the regime of finite doping, the RKKY interaction would take very larger values than undoped situation and oscillate, however this oscillation doesn't occur with a constant period which is originated from complicated form of the band structure. As mentioned in the previous section in Fig.\ref{fig.LDOS}, the density of states would have Van-Hove singularities at the edge of the band gap for $ V \neq 0  $ and as a result, as shown in Fig.\ref{fig.JEf} (c,d), the RKKY interaction would take very large values at very small doping (at the edge of conduction band) which is a result of Rashba-splitting in the band dispersion. Due to increasing of the DOS with energy for the intra-surface case, the RKKY interaction terms gradually increase with the Fermi energy. In the presence of Rashba splitting (panel c), all the terms first decrease by decrease of $k_{F2}$ and then after the critical Fermi energy $\varepsilon_F=\sqrt{V^2+\Delta^2}$ in which $k_{F2}$ becomes zero, they increase. The change in the Fermi wavevector $k_{F2}$ is proportional to a change in the electron's density on the top surface which justifies this behavior.

For the inter-surface case, the dominant parameter for controlling the RKKY interaction between two impurities located on different surfaces is not only the DOS on the top and bottom surfaces, but also the inter-surface hybridization of the surface states. In this case, as seen in Fig.\ref{fig.JEf} (b,d), the RKKY interaction terms decrease with respect to the energy. This can be described by the weight factors explained in Sec.\ref{Hamil} which say that at higher energies, the surface states are not hybridized any more and they will be purely localized on the top or the bottom surface which results in weakening of the inter-surface interaction.

Furthermore, as one can see in Fig. \ref{fig.JEf} (b), for $ V=0 $, there is only the Heisenberg interaction for the inter-surface case and other terms are exactly zero. In this case, the band dispersions belonging to different spin helical states wont split. This property together with the form of the tunnelling between surfaces, $ \Delta $, which does not couple different spins, make the RKKY interaction to be isotropic collinear.

\begin{figure}[H]
  \includegraphics[width=1\linewidth]{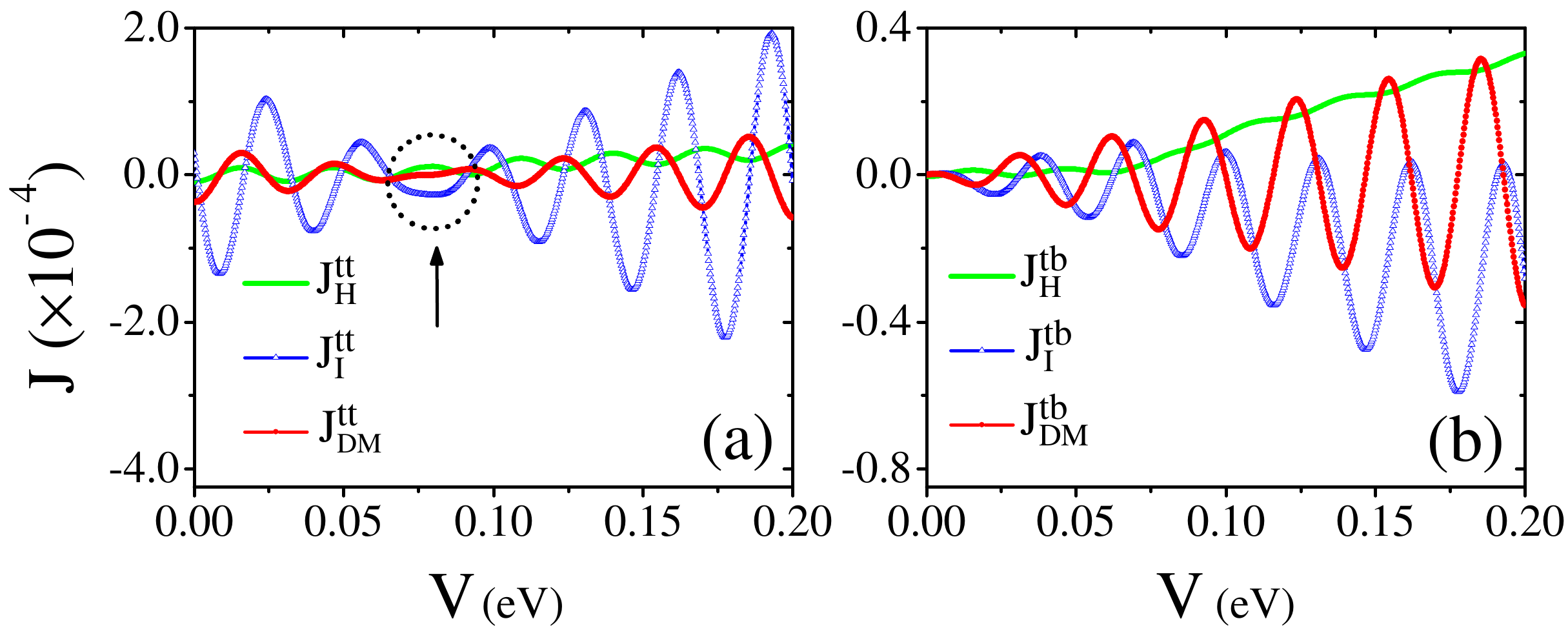}\\
  \caption{\small{(Color online) The RKKY interaction terms ($J^{\alpha \beta}_i$ for $ i=H, I, DM $), scaled by $ (\frac{J_c}{\hbar^2 v_F^2 \Omega_{BZ}})^2$, as a function of voltage. Here we set $R=30$ nm, $\Delta=0.035$ eV,  $\varepsilon_F=0.085$ eV and $v_{F}=4.48\times 10^5 \frac{m}{s}$. Panels (a) and (b) refer to intra-surface and inter-surface cases respectively.}}\label{fig.JV}
\end{figure}

To see the effect of Rashba splitting on the RKKY interaction, in Fig.\ref{fig.JV}, the behaviour of all RKKY interaction terms for intra- (panel a) and inter- (panel b) surface pairing are shown with respect to $V$. Fixing the chemical potential and changing the biased potential, one can tune the Fermi wave-vectors together with DOS and as a result tune the RKKY interaction. Tuning the magnetic properties of materials with electric field is so desirable for spintronic technologies.\cite{Ohno} For the intra-surface pairing depicted in Fig. \ref{fig.JV} (a), the RKKY interaction drops by decrease of $k_{F2}$  and then after the critical biased voltage $V=\sqrt{\varepsilon_F^2-\Delta^2}$ in which $k_{F2}=0$, it increases. At this critical voltage, the density of electrons on the top surface in which mediate the RKKY interaction becomes nearly zero and that's why the RKKY strength has its minimum in the dashed circle. The quenching the RKKY interaction terms in the critical voltage has the same root as its quench in the special Fermi energy shown in Fig. \ref{fig.JEf} (c). This transition in the RKKY interaction behaviour has been pointed out by an arrow and black dashed circle in the figure. For the inter-surface case both $k_F$s play a role and regardless of oscillations the interaction increases.

\subsection{Van-Vleck interaction}
The RKKY interaction refers to indirect exchange interaction via conduction's electrons which occurs in the metallic phase of systems. However, looking at Eq.\eqref{eq:8}, one can see the RKKY interaction is originated from all energies lower than the Fermi energy $\varepsilon_F$ as well, so it would have non-zero value even at zero chemical doping $\varepsilon_F=0$. Although, in this regime the indirect exchange interaction known as Van-Vleck interaction, is much weaker than the RKKY interaction, it can affect magnetic phases of materials.\cite{prl115}
The $zz$ component of the Van-Vleck interaction (related to $\chi_{zz}$) has been studied in TI thin films \cite{prl115} to describe the Ferromagnetic phase in QAHE experiment. Here we investigate all terms of this interaction and its tunability with the biased potential $V$.

\begin{figure}[H]
  \includegraphics[width=1\linewidth]{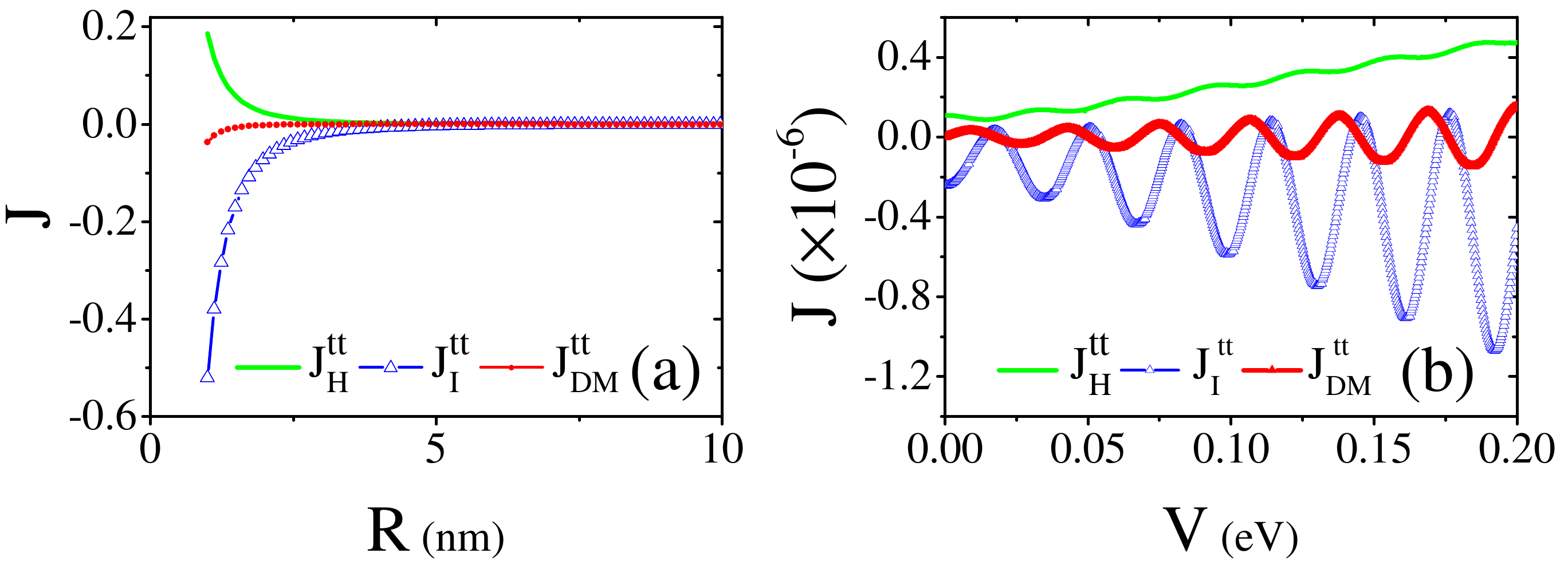}\\
  \caption{\small{(Color online) The Van-vleck interaction terms ($J^{\alpha \beta}_i$ for $ i=H, I, DM $), scaled by $ (\frac{J_c}{\hbar^2 v_F^2 \Omega_{BZ}})^2$, as a function of (a) distance, (b) voltage. Here we set $\Delta=0.035$ eV,  $\varepsilon_F=0.0$ eV and $v_{F}=4.48\times 10^5 \frac{m}{s}$ and for (a) $V=0.02$eV and (b) $R=30$ nm. }}\label{fig.JVan}
\end{figure}

Figure \ref{fig.JVan} (a) shows the Van-Vleck interaction with respect to distance R. As shown in this figure, all the interaction terms falls off very rapidly and becomes zero after $R \sim 3$nm. Panel (b) shows how this interaction is also oscillating with respect to the biased electric potential $V$. Moreover, it is obvious from this figure that $V$ make the Van-Vleck interaction stronger.

\section{Summary and Conclusion}\label{conclusion}
In summary, we have investigated the effect of Rashba-type band splitting on the RKKY interaction in topological insulator thin films. We explored the RKKY interactions for two different situations of magnetic impurities separately, namely inter-surface pairing and intra-surface pairing where we reported completely different behaviors. We describe this diversity by mapping the density of states onto the band dispersion and finding the share of each surface on the band dispersion.
We have shown how the RKKY interaction in the Rashba materials have a strong direction-dependency (spatial anisotropy) when at least one of the impurities has a spin component parallel to the plane. In addition to the conventional RKKY interaction terms mentioned in the Rashba materials, namely Heisenberg-like, Ising-like and DM-like, we found another term of the spin-susceptibility tensor which in contrast to the DM term is a symmetric interaction. We also investigate the RKKY interaction at zero doping where the chemical potential lies within the gap of the TI thin film, (usually known as the Van-Vleck mechanism). This can shed a light on solving the problem of QAHE which has been done experimentally at zero chemical doping. Furthermore, we show that how the Rashba splitting makes a Van-hove singularity in the band dispersion at the band edges giving rise to large values of the RKKY interaction. So by a small value of the chemical doping, the RKKY interaction can be extremely modified.

\section*{Acknowledgment }
F.P thanks Manuel Pereiro and Alireza Qaiumzadeh for useful discussions and M.Sh. thanks Saeed Amiri for his supportive role in the paper preparation's progress and also acknowledges "institute for research in fundamental sciences" for their hospitality while the last parts of this paper were preparing. H.C. thanks the International Center for Theoretical Physics (ICTP) for their hospitality and support during a visit
in which part of this work was done.

\appendix
\section{ِDetails of Green's function}\label{app}

Taking the integrals of Eq. \eqref{eq:9} according to the Fourier transformation, one can achieve the Green's function in real space. Using two-dimensional polar coordination in k-space, we have $ \exp(i \, \vec{\textbf{k}} \, \cdot \, \vec{\textbf{R}})= \exp(i \, k \, R \, \cos(\varphi_k-\varphi_R))  $ and so $G_0^{ret}(\varepsilon,\pm R)$ (Eq. \eqref{eq:10}) components would be obtained as the following

\begin{subequations}
\begin{align}
\label{eq:A1}
G_{tt}(\varepsilon,R)=&
	-2\pi \alpha \ \sum_{s=\pm} a_{-s}(\gamma-isV)K_{0}^s
	, \nonumber
\\
G'_{tt}(\varepsilon,R)=&
	-2\pi i \alpha \  \sum_{s=\pm} \frac{s a_{-s}}{\sqrt{\frac{-1}{(V+is\gamma)^2}}}K_{1}^s
	, \nonumber
\\
G_{tb}(\varepsilon,R)=&
	\pi i \alpha \ \frac{\Delta}{\gamma}\ \sum_{s=\pm} s (V+is\gamma)K_{0}^s
	, \nonumber
\\
G'_{tb}(\varepsilon,R)=&
	-\pi i \alpha \ \frac{\Delta}{\gamma}\ \sum_{s=\pm} \frac{s}{\sqrt{\frac{-1}{(V+is \gamma)^2}}} K_{1}^s
	, \nonumber
\\
G_{bb}(\varepsilon,R)=&
	-2\pi \alpha \ \sum_{s=\pm} a_s(\gamma-isV)K_0^s
	, \nonumber
\\
G'_{bb}(\varepsilon,R)=&
	-2\pi i \alpha \ \sum_{s=\pm} \frac{s a_s}{\sqrt{\frac{-1}{(V+is\gamma)^2}}}K_1^s
\end{align}
\end{subequations}
where, $\alpha=1/\hbar^2 v_F^2 \Omega_{BZ}$, $ \gamma=\sqrt{\Delta^2-\varepsilon^2} $ and for $ s=\pm $, $ a_{s}=\frac{1}{2}(\frac{\varepsilon}{\gamma}+s i) $ whereas $ K_{0/1}^s $ are the zeroth and first order of the  modified Bessel functions of the second kind as below
\begin{equation}\label{eq:B2}
\begin{split}
K_{0}^{s}=
	K_0\left(\frac{R}{\sqrt{-\frac{\hbar ^2 v_F^2}{(V+s i \gamma )^2}}}\right),
K_{1}^{s}=	K_1\left(\frac{R}{\sqrt{-\frac{\hbar ^2 v_F^2}{(V-s i \gamma )^2}}}\right).
\end{split}
\end{equation}

Re-writing the spin susceptibility as $\chi_{ij}^{\alpha \beta}=\frac{-1}{2\pi}Im\int_{-\infty}^{\varepsilon_F} d\varepsilon F_{ij}^{\alpha \beta}$, where $F_{ij}^{\alpha\beta}= Tr[\sigma_{i}\ G^{\alpha\beta}(r,r^{\prime},\varepsilon)\ \sigma_{j}\ G^{\beta\alpha}(r^{\prime},r,\varepsilon)]$ we can write the $F_{ij}^{tt(tb)}$s for the intra-surface case (t) and for the inter-surface case (tb) as:

\begin{equation}\label{eq:A3}
\begin{split}
F_{xx}^{tt(tb)}&
=2 \; (G_{tt(tb)}^2-G_{tt(tb)}^{'2} \; \cos(2 \varphi_R)),\\
F_{yy}^{tt(tb)}&
=2 \; (G_{tt(tb)}^2+G_{tt(tb)}^{'2} \; \cos(2 \varphi_R)),\\
F_{zz}^{tt(tb)}&
=2 \; (G_{tt(tb)}^2-G_{tt(tb)}^{'2}),\\
F_{xy}^{tt(tb)}&
=-2 \; G_{tt(tb)}^{'2} \; \sin(2 \varphi_R),\\
F_{xz}^{tt(tb)}&=-F_{zx}^{tt(tb)}
=4 \; G_{tt(tb)} \; G'_{tt(tb)} \; \cos(\varphi_R)),\\
F_{yz}^{tt(b)}&=-F_{zy}^{tt(b)}
=4 \; G_{tt(tb)} \; G'_{tt(tb)} \; \sin(\varphi_R).\\
\end{split}
\end{equation}
which after integration, it gives us the RKKY interaction terms. Introducing new spinors $\tilde{S}=(S_x\cos(\varphi),S_y\sin(\varphi),S_z)$ the RKKY interaction Eq.\eqref{eq:11} can be achieved easily.

\end{document}